\documentclass[%
reprint,
aps,
mph,
superscriptaddress,
nofootinbib,
nobibnotes,
amsmath,amssymb,
longbibliography
]{revtex4-1}

\usepackage{xcolor}

\usepackage{graphicx}
\usepackage{dcolumn}
\usepackage{bm}

\begin{document}
\newcommand{\eg}{{\it e.g.}}
\newcommand{\ie}{{\it i.e.}}

\preprint{APS/123-QED}

\title{Holographic multipartite entanglement from the upper bound of $n$-partite information}

\author{Xin-Xiang Ju}
\email{juxinxiang21@mails.ucas.ac.cn}
 \affiliation{School of Physical Sciences, University of Chinese Academy of Sciences, Zhongguancun east road 80, Beijing 100190, China}

\author{Wen-Bin Pan}%
 \email{panwb@ihep.ac.cn}
\affiliation{Institute of High Energy Physics, Chinese Academy of Sciences,\\19B Yuquan Road, Shijingshan District, Beijing 100049, China}%

\author{Ya-Wen Sun}
 \email{yawen.sun@ucas.ac.cn}
 \affiliation{School of Physical Sciences, University of Chinese Academy of Sciences, Zhongguancun east road 80, Beijing 100190, China}%
\affiliation{Kavli Institute for Theoretical Sciences, University of Chinese Academy of Sciences, Beijing 100049, China}%

\author{Yang Zhao}
\email{zhaoyang20a@mails.ucas.ac.cn}
\affiliation{School of Physical Sciences, University of Chinese Academy of Sciences, Zhongguancun east road 80, Beijing 100190, China}%

\date{\today}

\begin{abstract}
To analyze the holographic multipartite entanglement structure, we study the upper bound for holographic $n$-partite information $(-1)^n I_n$ that $n-1$ fixed boundary subregions participate together with an arbitrary region $E$. 
For $n=3$, we show that the upper bound of $-I_3$ is given by a quantity that we name the entanglement of state-constrained purification $EoSP(A:B)$. For $n\geq4$, we find that the upper bound of $I_n$ is finite in holographic CFT$_{1+1}$ but has UV divergences in higher dimensions, which reveals a fundamental difference in the entanglement structure in different dimensions. When $(-1)^n I_n$ reaches the information-theoretical upper bound, we argue that \( I_n \) fully accounts for multipartite global entanglement in these upper bound critical points, in contrast to usual cases where $I_n$ is not a perfect measure for multipartite entanglement. We further show that these results suggest that fewer-partite entanglement fully emerges from more-partite entanglement, and any $n-1$ distant regions are fully $n$-partite entangling in higher dimensions.   
\end{abstract}

\maketitle

\section{Introduction}
\noindent

As a strongly coupled quantum many-body system, holographic states \cite{Maldacena:1997re} are believed to have strong entanglement \cite{Ryu_2006} with a large amount of multi-partite entanglement \cite{Akers:2019gcv}. Various measures {and methods} have been proposed to detect the multi-partite entanglement structure \cite{Walter:2016lgl}, including {3-tangle \cite{greenberger2007goingbellstheorem,Coffman:1999jd,bengtsson2016brief}, multipartite reflected entropy \cite{Bao:2019zqc}, multipartite squashed entanglement \cite{Yang_2009,Yang_2008}}, combinations of signals \cite{Balasubramanian:2024ysu}, and holographic entropy cone \cite{Bao:2015bfa,Hubeny:2018trv,Hubeny:2018ijt,He:2019ttu,HernandezCuenca:2019wgh,He:2020xuo,Avis:2021xnz,Fadel:2021urx} etc. In this work, we investigate the multi-partite global entanglement in holographic states \cite{Ju:2023dzo} by studying the upper bound of the n-partite information \cite{Ju:2023tvo} $(-1)^n I_n$ that $n-1$ fixed boundary subregions participate together with an arbitrary subregion $E$.
\begin{equation}\label{I4def}
    {I}_{n}=\sum_{i}S_{A_i}-\sum_{i\neq j}S_{A_iA_j}+...+(-1)^{n+1}S_{A_1...A_n},
\end{equation}
serves as a measure of how much information is shared collectively among all $n$ subregions $A_i$. However, it is not a faithful measure as both quantum entanglement and classical correlations could contribute to $(-1)^nI_n$. The former is always positive \cite{Guo_2020} while the latter might be negative, making the sign of it indefinite \cite{Erdmenger:2017gdk}. 

Nevertheless, when $(-1)^nI_n$ reaches the information theoretical upper bound \cite{araki1970entropy}, only quantum entanglement contributes to it \cite{Shirokov_2017}.
In this work, we try to extract the holographic $n$-partite global entanglement that $n-1$ regions participate by analyzing the upper bound of $(-1)^n I_n$ with the $n$-th region arbitrarily chosen. $n$-partite global entanglement is the kind of multipartite entanglement which could exist when all $m$-partite ($m<n$) entanglement among subsystems vanishes.
In this work, we will start from $I_3$. Given distant regions $A$ and $B$ we develop a method to find the region $E$ that maximizes $-I_3(A:B:E)$, at which {bipartite entanglement vanishes between any two subsystems and the entanglement between $A$ and $BE$ reaches the maximum value}. Therefore, though $-I_3$ is in general not a good measure for tripartite entanglement, in the maximum configuration, it characterizes fully tripartite global entanglement. We will give a general formula for the upper bound value of $-I_3$. We will also generalize our method to find the upper bound of $I_4$, whose result reveals the fundamental difference of four-partite global entanglement structure in different dimensions.

\section{The upper bound of $-I_3$}

We analyze the upper bound of $-I_3(A:B:E)$ that two fixed regions $A$ and $B$ participate with any third region $E$ in holographic states, with
\begin{equation}\label{CMIformula}
\begin{aligned}
    -I_3(A:B:E)&=I(E:AB)-I(E:A)-I(E:B)\\
    &=I(A:B|E)-I(A:B),
\end{aligned}
\end{equation} where $I(A:B|E)=S_{AE}+S_{BE}-S_{ABE}-S_E$ is the conditional mutual information (CMI) between $A$ and $B$ under the condition $E$. In holography \cite{Ryu_2006,Ryu:2006ef,Headrick:2019eth}, $I_3$ is always negative due to the monogamy of {mutual information} \cite{Hayden_2013}.
\footnote{In \cite{Cui:2018dyq}, it is proposed that $I_3$ measures the pure state perfect-tensor-type four-partite entanglement, which is not the mixed state four-partite global entanglement we investigate here.}  
As $I(A:B)$ is fixed, finding region $E$ that maximizes $-I_3$ is equivalent to finding $E$ that maximizes $I(A:B|E)$. 

In quantum information theory, the information-theoretical upper bound of $-I_3$ is $2\min (S_A,S_B)$ due to the Araki-Lieb inequality \cite{araki1970entropy}. When the upper bound is saturated, taking $S_A\leq S_B$ without loss of generality, we need to have 
\begin{equation}\label{CMIsaturate}
    I(A:E)=I(A:B)=0,\,\, I(A:BE)=2S_A,
\end{equation}
meaning that $A$ does not participate in the bipartite entanglement with $E$ or $B$. $I(A:BE)=2S_A$, which is the information-theoretical upper bound and thus is fully quantum \cite{Shirokov_2017}, indicates that $A$ contributes all its d.o.fs participating in the tripartite global entanglement \cite{bengtsson2016brief} with $B$ and $E$. Therefore, at saturation of the upper bound $-I_3=2 S_A$, $I_3$ fully captures the tripartite entanglement in contrast to not being a faithful measure in general cases.

The existence of such a region $E$ that saturates the upper bound of $-I_3$ implies important features in the multipartite entanglement structure. Therefore, we seek the region $E$ that maximizes $-I_3$ and check if it could saturate the information-theoretical upper bound of $-I_3$ in holography in AdS$_3$/CFT$_2$. All the conclusions could be easily generalized to higher dimensions. In holography, $-I_3$ is an IR term without UV divergence because all UV divergences in the mutual information cancel out in $-I_3$. The only chance for $-I_3$ to approach its information-theoretical upper bound $2S_A$, which is UV divergent, is by making the number of intervals in $E$ approach infinity.
However, analyzing the maximum $I(A:B|E)$ for $E$ having infinitely many intervals is technically formidable. 

In this letter, we develop an elegant method to find the region $E$ that maximizes $-I_3$. Motivated by (\ref{CMIsaturate}), we propose and will prove later that this region $E$ should satisfy the following constraints on the connectivity of the entanglement wedges \cite{Czech:2012bh,Wall:2012uf,Headrick:2014cta} $EW(AE),EW(BE),EW(E)$, and $EW(ABE)$ as follows.
\begin{itemize}
    \item \textbf{I}. $EW(ABE)$ being totally connected.
    \item \textbf{II}. $EW(E)$ being totally disconnected.
    \item \textbf{III. disconnectivity condition:} $EW(AE)$ and $EW(BE)$ being disconnected {\ie, $I(A:E)=I(B:E)=0$}.
\end{itemize} These conditions greatly reduce the difficulty of obtaining the configuration with maximum $-I_3$ as we only need to pick the $E$ that has maximum $-I_3$ from all regions $E$ that satisfy these constraints and the expression of $-I_3$ gets greatly simplified under these conditions, too.
To prove these conditions, the core idea is to prove: \textit{given a configuration of $E$ that does not satisfy these conditions, there always exists another configuration of $E$ that satisfies these conditions, with $-I_3$ or equivalently $I(A:B|E)$ not less than the former one}. Let us prove these conditions one by one.

\textbf{I}. If there exists a single interval \( E_i \) where, in \( ABE \)'s entanglement wedge, \( E_i \) is disconnected—which means \( I(ABE\backslash E_i : E_i) = 0 \)—one can find
\begin{equation}\label{ABEcon}
\begin{aligned}
    I(A:B|E) &= S_{AE} + S_{BE} - S_E - S_{ABE} \\
             &= S_{AE\backslash E_i} + S_{E_i} + S_{BE\backslash E_i} + S_{E_i} \\
             &\quad- S_{E\backslash E_i} - S_{E_i} - S_{ABE\backslash E_i} - S_{E_i} \\
             &= I(A:B|E\backslash E_i),
\end{aligned}
\end{equation}
i.e., the CMI will be the same if we delete \( E_i \) from \( E \). We can perform this procedure repeatedly until all disconnected $E_i$ are deleted, so that \textbf{I} is satisfied. Therefore, we only need to consider configurations that satisfy \textbf{I}.

\textbf{II}. If  \( EW(E) \) is partially connected—let us say, \( E_i \) and \( E_{i+1} \) are connected {in $EW(E)$,}—we denote the gap region between \( E_i \) and \( E_{i+1} \) as \( G_{i} \). Then we can ``merge'' \( E_i \), \( G_{i} \), and \( E_{i+1} \) into a single interval to replace the former \( E_i \) and \( E_{i+1} \). Denoting the new region \( E \) as \( E_{\text{new}} \), we have
\begin{equation}
\begin{aligned}
    I(A:B|E)&=S_{AE}+S_{BE}-S_E-S_{ABE}    \\
            &=S_{AE_{\text{new}}}+S_{G_{i}}+S_{BE_{\text{new}}}+S_{G_{i}}\\
            &\quad-S_{E_{\text{new}}}-S_{G_{i}}-S_{ABE_{\text{new}}}-S_{G_{i}}\\
            &=I(A:B|E_{\text{new}}),
\end{aligned}
\end{equation}
i.e., the CMI will be the same if we merge \( E_i \), \( G_{i} \), and \( E_{i+1} \) together. Here we have used the fact that $E_{i}$ and $E_{i+1}$ must be connected in all entanglement wedges. We can perform this procedure repeatedly until all connected $E_i$ are merged together so that \textbf{II} is satisfied. Therefore, we only need to consider configurations that satisfy \textbf{II}.

\textbf{III}. If there exists a region $E_i$ which connects with $A$ in $EW(AE)$ or connects with $B$ in $EW(BE)$, we can split $E_i$ into three regions $E_{i1}$, $E_{i2}$, and the gap region $G_{i}$ between them, with $E_{i1}$ and $E_{i2}$ disconnected from $A$ in $EW(AE)$ or disconnected from $B$ in $EW(BE)$. During this procedure, CMI does not decrease. We can perform this procedure repeatedly until $I(A:E)=I(B:E)=0$, satisfying \textbf{III}. More detailed proof can be found in Appendix A.  

With these three constraints on the connectivity of the entanglement wedges proved, we can evaluate the upper bound of CMI. In the following, we calculate the maximum CMI in two cases: asymptotic AdS$_3$ and two-sided black hole geometry.

\begin{figure}[h]
\centering
\includegraphics[width=9cm]{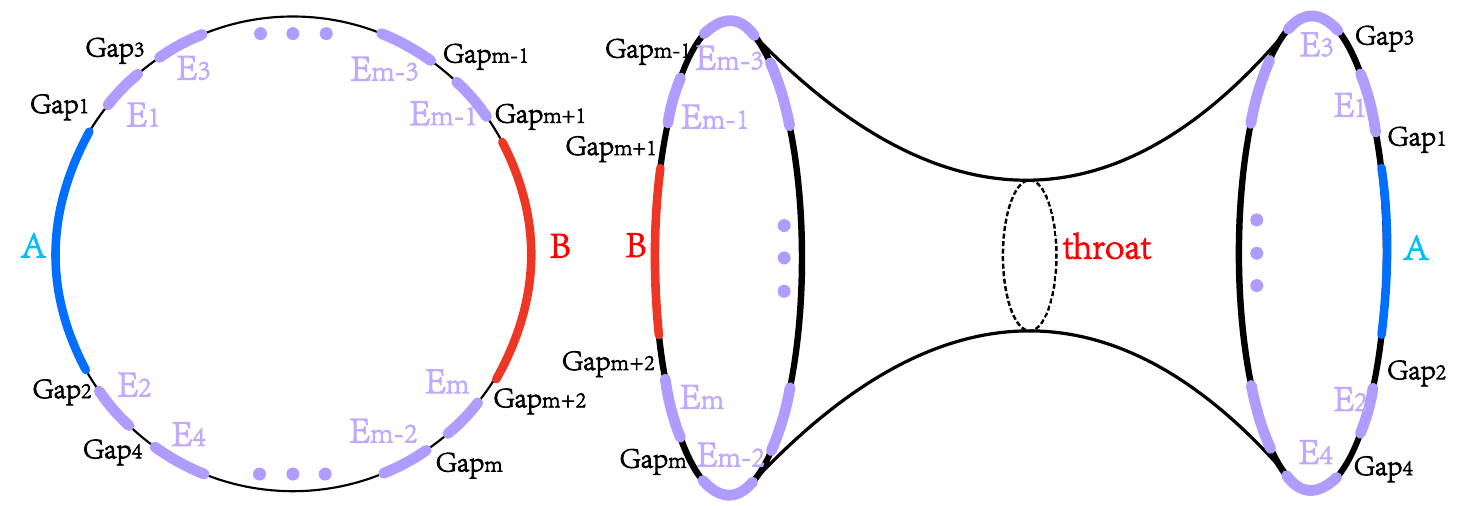}
\caption{Illustration for finding the maximum CMI configurations when $m$ intervals of $E$ could live anywhere that has no overlap with $A$ or $B$ for asymptotic AdS$_3$ (left) and the two-sided black hole (right). Regions \( A \), \( B \), and \( E \) are marked by blue, red, and purple intervals, respectively. The \( m + 2 \) gap regions between \( E_i \), \( A \), and \( B \) are labeled as \( \text{Gap}_i \).}
\label{GENCMI}
\end{figure}

We first consider the case when $A$ and $B$ are two intervals on the $S^1$ spatial boundary in asymptotic AdS$_3$. We assume that $E$ is a collection of $m$ intervals living in both the gap regions between $A$ and $B$. The left side of Figure \ref{GENCMI} depicts the configuration of such an $E$ in global coordinates. Using these connectivity conditions of the entanglement wedges, the formula of CMI reduces to
\begin{equation}\label{generalCMI}
\begin{aligned}
    I(A:B|E) &= S_{AE} + S_{BE} - S_E - S_{ABE} \\
             &= S_A + S_E + S_B + S_E - S_E - S_{ABE} \\
             &= S_A + S_B + \sum_{i=1}^m S_{E_i} - \sum_{i=1}^{m+2} S_{\text{G}_i},
\end{aligned}
\end{equation} where $S_{G_i}$ denotes the entanglement entropy of the gap region $\text{Gap}_i$. In the second and third lines, \textbf{III} and \textbf{I}, \textbf{II} are used, respectively.

The vanishing of mutual information $I(E:A)$ and $I(E:B)$ in \textbf{III} implies that the surface homologous to $BE$ has minimal area for the disconnected configuration rather than the connected configuration
\begin{equation}\label{AEdiscon}
    \begin{aligned}
    S_B + \sum_{i=1}^m S_{E_i} &\leq \sum_{i=1}^{m+2} S_{\text{G}_i} - S_{\text{G}_{m+1}} - S_{\text{G}_{m+2}} + S_{\text{G}_{m+1} A \text{G}_{m+2}}. 
    \end{aligned}
\end{equation}
 The left-hand side of this inequality corresponds to the area of the surface homologous to $BE$ which has a disconnected configuration, while the right-hand side corresponds to the area of the surface homologous to $BE$ which has a connected configuration. This inequality gives $\sum_{i=1}^m S_{E_i} - \sum_{i=1}^{m+2} S_{\text{G}_i}$ an upper bound, which in turn gives CMI in formula (\ref{generalCMI}) an upper bound as follows.
\begin{equation}
    I(A:B|E) \leq S_{\text{G}_{m+1} A \text{G}_{m+2}} + S_A - S_{\text{G}_{m+1}} - S_{\text{G}_{m+2}}.
\end{equation}
Assuming $S_{A}\leq S_B$ without loss of generality, this inequality can be saturated when $BE$ reaches its phase transition point from the disconnected configuration to the connected configuration. When $m$ approaches infinity, the length of every single gap region goes to zero so that $S_{\text{G}_{i}}$ tends to zero and $S_{\text{G}_{m+1} A \text{G}_{m+2}}$ tends to $S_A$. Thus, we have that $I(A:B|E)$ approaches $2S_A$, its information-theoretical upper bound at last.

Therefore, for small distant subsystems $A$ and $B$, the d.o.fs in $A$ all participate in the tripartite entanglement with $B$ and another carefully chosen region $E$. We can make the following bold statement.
\begin{itemize}
    \item No Bell pairs exist in holographic states; all bipartite entanglement emerges from tripartite entanglement; any two distant small subsystems are highly tripartite entangling with another system.
\end{itemize}

We then analyze the upper bound of $-I_3$ in the two-sided black hole geometry \cite{Maldacena:2013xja,Susskind:2014yaa}, as shown on the right side of Figure \ref{GENCMI}. When $A$ and $B$ are on the same side of the black hole, the upper bound is the same as in asymptotic AdS$_3$ through similar analysis. However, when \( A \) and \( B \) live in two different boundary CFTs, as we will see, the upper bound will be much less with no UV divergence due to the long-range entanglement \cite{Ju:2024xcn} involved. It is worth noting that in this case, \textbf{I}, \textbf{II}, and \textbf{III} are still valid as no specific background geometry is stipulated when proving them. As a result, (\ref{generalCMI}) is still valid. The key step is still calculating the maximum value of \( \sum_{i=1}^m S_{E_i} - \sum_{i=1}^{m+2} S_{\text{G}_i} \), which is determined by the disconnectivity condition of \( EW(AE) \), giving a different inequality in this case:
\begin{equation}
\begin{aligned}
 S_A + \sum_{i=1}^{m} S_{E_i} \leq \sum_{i=1}^{{m}/{2}+1} S_{\text{G}_i} + \sum_{i={m}/{2}+1}^{m} S_{E_i} + S_{\text{throat}}.
\end{aligned}
\end{equation}
The right hand side corresponds to a partially connected configuration of $EW(AE)$ where intervals of \( E \) on the same boundary as region \( A \) (right boundary in Figure \ref{GENCMI}) connect with \( A \), while the intervals of \( E \) on the other boundary (left boundary in Figure \ref{GENCMI}) disconnect with \( A \).
Using the symmetry of exchanging \( A \) and \( B \), we have
\begin{equation}
     S_B + \sum_{i=1}^{m} S_{E_i} \leq \sum_{i={m}/{2}+2}^{{m}+2} S_{\text{G}_i} + \sum_{i=1}^{{m}/{2}} S_{E_i} + S_{\text{throat}}.
\end{equation}
Adding these two inequalities together, we have
\begin{equation}
\begin{aligned}
    S_A + S_B + \sum_{i=1}^{m} S_{E_i} - \sum_{i=1}^{m+2} S_{\text{G}_i} &\leq 2S_{\text{throat}}, \\
  \ie, \quad  I(A:B|E) &\leq 2S_{\text{throat}}.
\end{aligned}
\end{equation}
Note that this inequality can be saturated for sufficiently large but finite $m$.

This is a surprising result, as we always take the throat as a symbol of bipartite entanglement between two CFT boundaries. However, we found that the throat area contributes to the tripartite entanglement between arbitrarily chosen sufficiently small regions \( A \) and \( B \) with another region \( E \). Compared with the $AdS_3$ result, this indicates the long-range nature in the tripartite global entanglement in $ABE$.

After obtaining the upper bound of $-I_3$ in the AdS$_3$ and the two-sided black hole cases, we give a general formula for the value of the upper bound of $-I_3$ in the following. In the most general case, e.g. when $A$ or $B$ has multiple intervals or in higher dimensions, formula (\ref{generalCMI}) must still be satisfied. The only difference is that \textbf{III} will lead to different constraints on \(\sum_{i=1}^m S_{E_i} - \sum_{i=1}^{m+2} S_{\text{G}_i}\) in general as follows
\begin{equation}
\begin{aligned}
S_A + \sum_{i=1}^m S_{E_i} &\leq \sum_{i \in D} S_{\text{G}_i} + \sum_{i \in D^c} S_{E_i} + S_D,
\end{aligned}
\end{equation}
where the left-hand side still corresponds to the area of the disconnected configuration of $EW(AE)$. We define a region $D\supseteq A$, $D^c\supseteq B$, and $E$ is divided into two parts: one in $D$ and one in $D^c$. The right-hand side corresponds to a partially connected configuration of $EW(AE)$ where, inside region $D$, $AE$ is totally connected, while inside region $D^c$, $AE$ is disconnected.

According to the exchange symmetry of $A$ and $B$, we can substitute $A$ and $D$ with $B$ and $D^c$, respectively. We have
\begin{equation}
S_B + \sum_{i=1}^m S_{E_i} \leq \sum_{i \in D^c} S_{\text{G}_i} + \sum_{i \in D} S_{E_i} + S_{D^c},
\end{equation}
Adding these two inequalities together, we have
\begin{equation}\label{FinalCMI}
I(A:B|E) \leq 2S_D \quad\quad (D \supseteq A, \ D^c \supseteq B).
\end{equation}
To make this upper bound as tight as possible, we need $S_D$ to be the minimal surface dividing $EW(A)$ from $EW(B)$. As before, this upper bound can be infinitely approached when $m \to \infty$. This upper bound is the most general formula. We can check that in the two-sided black hole, $D$ is the entire side which contains $A$, and in the global AdS$_3$/CFT$_2$, $D$ is either $A$ or $B^c$, whichever has the smaller entanglement entropy, both giving the previous results.

In AdS$_4$/CFT$_3$, it can be proved that this general formula (\ref{FinalCMI}) is still valid; \ie, CMI approches $2\min\{S_D\}$ when $m$ goes to infinity. Moreover, when $A$ (or $B$) is a concave region, the minimal surface that divides $A$ and $B$ will be the RT surface of the convex hull of $A$ (or $B$), instead of the RT surface of $A$ (or $B$) itself. \textit{This reveals the relationship between the convexity of the shape of the boundary region and the tripartite entanglement it participates in.} Specifically, the UV degrees of freedom of region $A$ near the concave part (the part which is near $\partial A$ but not near the boundary of the convex hull of $A$) cannot fully participate in the tripartite entanglement with a region outside the convex hull of $A$.

We would like to impart a physical meaning to the area of the minimal surface $S_D$ in the context of holographic quantum information theory. The definition of the RT surface of $D$ is very similar to the definition of the entanglement wedge cross section (EWCS). In \cite{Takayanagi:2017knl}, the EWCS between $A$ and $B$ is defined as the minimal surface that divides $EW(AB)$ into two parts, with its area being conjectured to be the entanglement of purification (EoP) \cite{Terhal_2002,Bagchi_2015} between $A$ and $B$. 

When $EW(AB)$ is disconnected, no surface is needed to divide it into two parts. The area of the EWCS vanishes in this case, as does the EoP when $I(A:B) = 0$ to leading order in $\frac{1}{G_N}$.
However, in our case, $S_D$ does not vanish when $I(A:B) = 0$. Instead, it is a large value. Therefore, it is not the traditional EoP, but rather the entanglement of state-constrained purification (EoSP) that we define as follows:
\begin{equation}
\begin{aligned}
&\text{EoSP}(A:B) = \min (S_{AA'}=S_{BB'}), \quad \\
&\text{where } A'B'=(AB)^c \text{ are boundary subregions}.
\end{aligned}
\end{equation}
The only difference between EoP and EoSP is that $A'$ and $B'$ are constrained to be boundary subregions in this holographic state instead of arbitrary extensions in quantum information theory. This constraint makes EoSP not the optimal choice from the viewpoint of EoP, resulting in it being significantly larger than EoP. We can easily prove that the value of EoSP we define is exactly $S_D$, as the region $AA'$ is region $D$.

Finally, we find the general tight upper bound of CMI
\begin{equation}\label{CMIEoSP}
\sup_{E} I(A:B|E) = 2\text{EoSP}(A:B),
\end{equation}
which can be saturated in general.

\section{The upper bound of $I_4$}

In this section, we analyze the upper bound of $I_4(A:B:C:E)$ with $A$, $B$, $C$ fixed and $E$ arbitrarily chosen in holography. In quantum information theory, the amount of four-partite entanglement is independent of tripartite and bipartite entanglement among subsystems  \cite{Ananth_2015}, and $I_4$ reaches its upper bound when tripartite entanglement vanishes. 
We could also stipulate the connectivity of the entanglement wedges as we did in the last section. The only difference is that the disconnectivity condition \textbf{III} should be modified as 
\begin{equation}
    I(E:AB)=I(E:BC)=I(E:AC)=0.
\end{equation}
The proof of this new constraint is tedious, but basically the same as we did for $I_3$, which can be found in Appendix B. We only need to consider $E$ as a combination of intervals which satisfy those constraints to evaluate the upper bound of $I_4$, which would be greatly simplified under these constraints
\begin{equation}\label{I4indiscon}
\begin{aligned}
    I_4(E:A:B:C)&=S_{ABC}+S_E-S_{ABCE}\\
    &=S_{ABC}+\sum_{i} S_{E_i} - \sum_{i} S_{\text{G}_i}.
\end{aligned}
\end{equation}
As a result, calculating the upper bound of \( I_4 \) is still a problem of calculating the maximum value of \( \sum_{i} S_{E_i} - \sum_{i} S_{\text{G}_i} \), and the disconnectivity condition constrains its value from being too large. 

\begin{figure}[h]
\centering
\includegraphics[width=8.5cm]{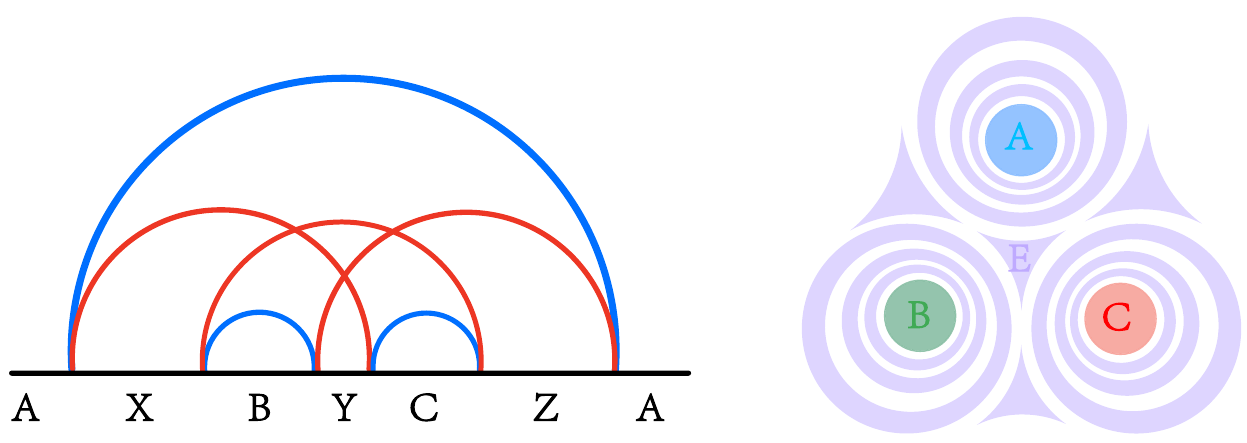}
\caption{\( I_4 \) in holographic \(\text{CFT}_{1+1}\) and \(\text{CFT}_{2+1}\). \( A \), \( B \), and \( C \) are intervals (left) and disks (right), respectively.}
\label{I4calculation}
\end{figure}

Let us start from AdS$_3$/CFT$_2$ as shown on the left side in Figure \ref{I4calculation}. \( I(E:AB)=0 \) stipulates that
\begin{equation}
    S_{AB}+\sum_{E_i\in X} S_{E_i} \leq \sum_{ \text{G}_i \in X} S_{\text{G}_i} + S_{YCZ}.
\end{equation}
This inequality prevents the intervals inside $X$ from connecting with $AB$ in $EW(ABE)$, which would violate the disconnectivity condition.
Using \( I(E:AC)=I(E:BC)=0 \), we can get the other two inequalities as follows:
\begin{equation}
    S_{AC}+\sum_{E_i\subset Z} S_{E_i} \leq \sum_{ \text{G}_i \subset Z} S_{\text{G}_i} + S_{XBY},
\end{equation}
\begin{equation}
    S_{BC}+\sum_{E_i\subset Y} S_{E_i} \leq \sum_{ \text{G}_i \subset Y} S_{\text{G}_i} + S_{ZAX}.
\end{equation}
Adding these three inequalities together, combined with formula (\ref{I4disconnect}), we can get
\begin{equation}
\begin{aligned}
    I_4 \leq \,\,&S_{YCZ} + S_{XBY} + S_{ZAX}  \\&- S_{AB} - S_{BC} - S_{AC} + S_{ABC},
\end{aligned}
\end{equation}
Finally, we get the upper bound of \( I_4 \). Specifically, on the left side of Figure \ref{I4calculation}, this upper bound is the summation of the length of red curves minus the summation of the length of blue curves plus $I_3(A:B:C)$. We can find that the UV divergent terms of the red curves and the blue curves cancel out. As $I_3(A:B:C)$ is an IR term, it will only finitely affect the value of \( I_4 \). As a consequence, the upper bound of \( I_4 \) is definitely a finite value without UV divergence. This result can be generalized to general AdS$_3$/CFT$_2$ cases with $A$, $B$, and $C$ each being multiple intervals.

Let us calculate \( I_4 \) in higher-dimensional holography. As shown on the right side of Figure \ref{I4calculation}, \( E \) is chosen as strips (annuli) which surround regions \( A \), \( B \), and \( C \). Formula (\ref{I4indiscon}) is still valid. The only difference is the constraint on \( \sum_{i} S_{E_i} - \sum_{i} S_{\text{G}_i} \) imposed by the disconnectivity condition, which is
\begin{equation}\label{I4disconstraint}
    S_{AB}+\sum_{i} S_{E_i} \leq \sum_{i} S_{\text{G}_i} - S_{\text{G}_m} + S_{C \text{G}_m},
\end{equation}
where \( C \text{G}_m \) is the gap adjacent to \( C \), which will be infinitely thin when \( m \) tends to infinity. Rewriting the last inequality as
\begin{equation}
\begin{aligned}
    S_{ABC}+\sum_{i} S_{E_i} - \sum_{i} S_{\text{G}_i} &\leq S_{C} - S_{AB} + S_{ABC}, \\
    \ie, \quad I_4 &\leq 2S_C - I(AB:C).
\end{aligned}
\end{equation}
Here, we assume that $2S_C - I(AB:C)$ is minimal among \{$2S_C - I(AB:C),2S_B - I(AC:B),2S_A - I(BC:A)$\} without loss of generality. This upper bound can be saturated in a way similar to the last section. 

At last, we conclude that \textit{given $A,B,C$ fixed and $E$ arbitrarily chosen, the upper bound of $I_4$ is finite in AdS$_3$/CFT$_2$ but infinite in higher-dimensional holography.} This reveals the fundamental difference between the multipartite entanglement structure in low-dimensional holography and higher-dimensional holography. In higher-dimensional holographic CFT, any three distant local d.o.f.s are maximally participating in four-partite entanglement, while in AdS$_3$/CFT$_2$, the four-partite entanglement among them is sparse.

To further investigate the more-partite global entanglement structure that $n-1$ small distant regions participate, we can analyze the upper bound of $(-1)^nI_{n>4}$ under a generalized disconnectivity condition. Fortunately, in higher dimensional holography, we can always construct a configuration of the $n$-th region such that, $I_n$ in this configuration reaches $2S_A$ while any $m$-partite entanglement for $m$ of the subregions with $m<n$ vanishes! \textit{This implies that these subregions fully participate in the $n$-partite global entanglement, where all $m$-partite entanglement among $m$-partitions of these $n$ subregions arises from it} \cite{Ju:inprogress}.
However, in AdS$_3$/CFT$_2$, $I_{n>4}$ is a summation of $I_4$, and as $I_4$ has finite upper and lower bounds, the upper bound of $I_{n>4}$ will always be finite, \ie, $n$-partite global entanglement is sparse among these subregions.

\section{Conclusion}

In this paper, we have investigated the tripartite entanglement structure in holographic states by evaluating the upper bound of $-I_3(A:B:E)$ that $A$ and $B$ participate. The general formula for the upper bound is $2EoSP(A:B)-I(A:B)$ which reaches the information theoretical upper bound in a wide class of $A$ and $B$. Due to the properties at this upper bound, $-I_3$ fully captures the tripartite global entanglement. As we have explained, this result reveals that all bipartite entanglement emerges from the tripartite entanglement in holographic states, and this conclusion can be generalized to $n\geq 4$-partite global entanglement in higher-dimensional holography. However, the upper bound of $(-1)^n I_{n\geq 4}$ is finite in AdS$_3$/CFT$_2$, which reveals the fundamental difference in multipartite entanglement structure in different dimensions.

\section*{Acknowledgement}

This work was supported by Project 12347183, 12035016 and 12275275 supported by the National Natural Science Foundation of China.
It is also supported by Beijing Natural Science Foundation under Grant No. 1222031.
\bibliography{apssamp}

\appendix

\section{Proof of the disconnectivity condition for $I_3$}

In this appendix, we prove the disconnectivity condition as follows.
\textit{Given a configuration $E$ with a non-vanishing $I(A:E)$ or $I(B:E)$, there always exists a disconnected configuration of $E$ with vanishing $I(A:E)$ and $I(B:E)$ whose CMI $I(A:B|E)$ is not less than the former one.}

Note that we do not demand that the disconnected configuration has the same $m$ as the connected configuration. One can choose a configuration with $m$ many times larger than that of the connected configuration, as long as its CMI is not less. 

\begin{figure*}
\centering
\includegraphics[width=12cm]{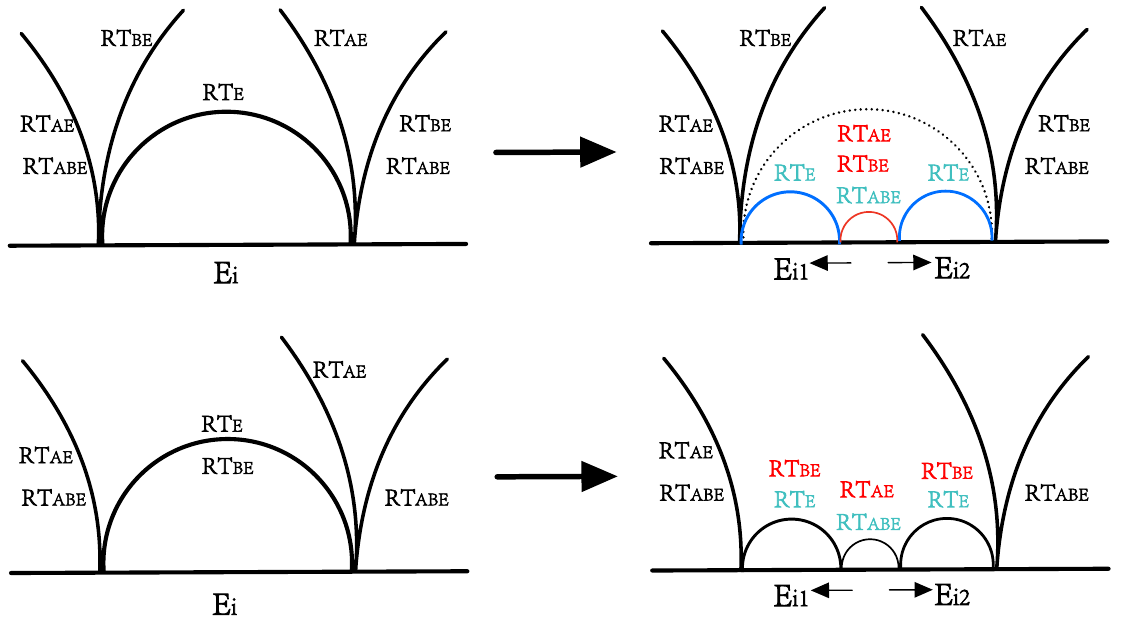}
 \caption{Proof of the weaker form disconnectivity condition. Each RT surface of region $M$ is labeled as $RT_{M}$. Only the surfaces whose area will change under modifying the endpoints of the gap are marked by blue and red, which represent their signature in formula (\ref{CMIformula}).}
\label{CMIDISPRO}
\end{figure*}

Let us prove it in the general situation. There are four steps to prove this theorem. According to Figure \ref{CMIDISPRO}, we start from {one of the intervals} of $E$, written as $E_i$, whose entanglement wedge is connected with both $A$ and $B$ in the configuration of $EW(AE)$ and $EW(BE)$. We split $E_i$ from the middle into two subregions $E_{i1}$ and $E_{i2}$ by adding a gap inside while preserving the outer boundary of $E_i$ unchanged. When the gap is very small, the $E_{i1}$ and $E_{i2}$ are fully connected in $EW(E)$, and as we have argued in the proof of condition \textbf{II}, the value of CMI does not change after opening this small gap.

The second step: when we continuously enlarge the gap, the first phase transition of the entanglement wedge occurs, which makes $E_{i1}$ and $E_{i2}$ disconnected in $EW(E)$, while preserving the connectivity of $E_{i1}$ and $E_{i2}$ in $EW(AE)$ and $EW(BE)$. This is shown in Figure \ref{CMIDISPRO}.
 
The third step: we continuously enlarge the gap inside $E$. One can find that $S_{E_{i1}}$ and $S_{E_{i2}}$ both decrease, while the entropy of the gap region in-between $S_{gap}$ increases. These three quantities are the only ones that affect the value of CMI because the outer boundary of $E_i$ remains unchanged. The sign of them inside {$I(A:B|E)$} is drawn as the different colors of those RT surfaces in Figure \ref{CMIDISPRO}; red represents positive and blue represents negative. It is easy to observe that when we enlarge the gap, CMI increases until another phase transition occurs. In principle, the phase transition could occur on one of $E_{i1}$ or $E_{i2}$, \ie $EW(AE)$ suddenly becomes disconnected between $A$ and $E_{i1(2)}$ while remaining the connectivity between $A$ and $E_{i2(1)}$ ({Here, we assume that $E_{i1}$ and $E_{i2}$ become disconnected in $EW(AE)$ before $E_{i1}$ and $E_{i2}$ become disconnected in $EW(BE)$, without loss of generality.}). However, we can make the phase transition occur on $E_{i1}$ and $E_{i2}$ simultaneously by adjusting the middle point of this gap. {Note that if the middle point of the gap region between $E_{i1}$ and $E_{i2}$ is far left inside the original region $E$, $E_{i1}$ disconnects with $A$ first; otherwise, if the middle point is far right inside the original region $E$, $E_{i2}$ disconnects with $A$ first. Therefore, there must exist a fine-tuned middle point which makes the phase transition occur on $E_{i1}$ and $E_{i2}$ simultaneously. After the phase transition, $E_{i1}$ and $E_{i2}$ disconnect with $A$ in $EW(AE)$.} Remember that when we enlarge the gap, the CMI increases, so we prove that:
\textit{for a configuration whose intervals of $E$ connect with both $A$ and $B$ inside $EW(AE)$ and $EW(BE)$ respectively, one can always find another configuration whose intervals connect to at most one of $A$ and $B$ inside $EW(AE)$ and $EW(BE)$, with larger CMI.}

After performing the third step for all the intervals of $E$, we are left with a configuration where all intervals of $E$ only connect with $A$ in $EW(AE)$ or only connect with $B$ in $EW(BE)$. The fourth step is to consider an interval $E_j$ which is only connected with $B$ in $EW(BE)$, and find another disconnected configuration with CMI not less than it. To achieve this final goal, we have to split $E_j$ again. In the second line of Figure \ref{CMIDISPRO}, this process is illustrated. The only difference is that the CMI will not change no matter how we enlarge the gap, as long as $BE_{j1}$ and $BE_{j2}$ are connected in $EW(BE)$. Again, we can make the phase transition occur between $E_{j1}$ and $B$ and between $E_{j2}$ and $B$ simultaneously. At the end, both $E_{j1}$ and $E_{j2}$ are disconnected from $B$. Combining with the last three steps, in the end, we find a disconnected configuration with CMI {not less than} that of the connected configuration, and the disconnectivity condition is proven. {Note that this proof is valid for various geometries and in higher-dimensional cases. When \( E_i \) is a disk, the ``gap'' inside region \( E_i \) would be a disk; when \( E_i \) is an annulus, the gap would be an annulus that splits \( E_i \) into two annuli, etc.}

\section{Proof of the disconnectivity condition for $I_4$}

The disconnectivity condition in \( I_4 \) is a stronger generalization of that in \( I_3 \), which states that only analyzing the disconnected case which satisfies
\begin{equation}\label{I4disconnect}
    I(E:AB) = I(E:AC) = I(E:BC) = 0,
\end{equation}
is enough to evaluate the upper bound of \( I_4 \). In other words, for a connected configuration of \( E \) which does not satisfy formula (\ref{I4disconnect}), there always exists a disconnected configuration \( E \) whose \( I_4 \) is not less.

\begin{figure*}
\centering
\includegraphics[width=15cm]{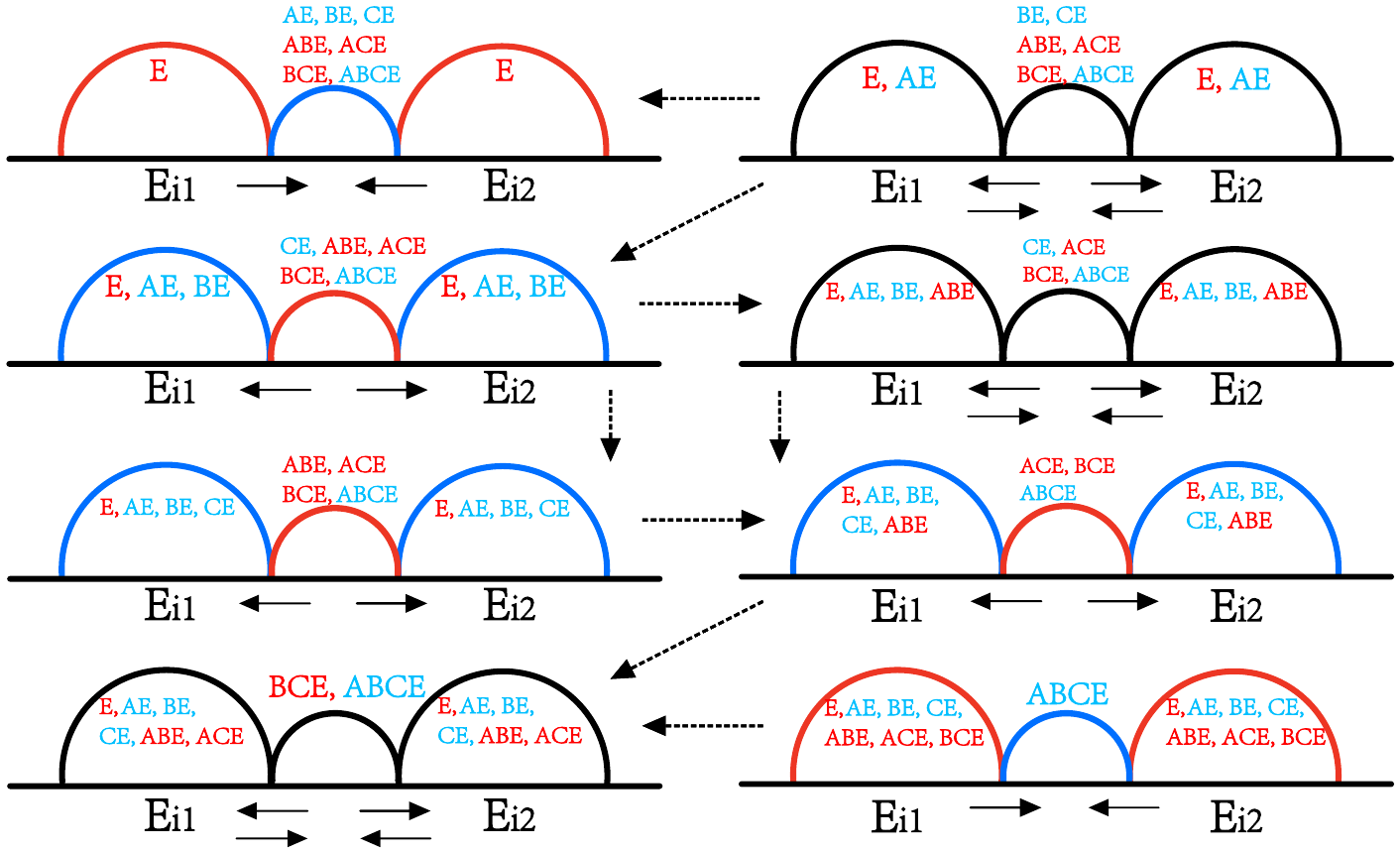}
\caption{Proof of the \( I_4 \) version of the disconnectivity condition.}
\label{DISI4}
\end{figure*}

To prove this theorem, we split the interval \( E_i \) which connects to at least one of \( A \), \( B \), and \( C \) in \( EW(ABE) \) or \( EW(BCE) \) or \( EW(ACE) \) into two pieces, \( E_{i1} \) and \( E_{i2} \), as we did in the last section. Figure \ref{DISI4} presents this whole procedure.

In this figure, each of the eight diagrams presents a case where interval \( E_i \) is split into two halves \( E_{i1} \) and \( E_{i2} \). When the position of the endpoints of the gap is modified, only the area of three minimal surfaces (semicircles in the figure) changes to affect \( I_4 \). According to formula (\ref{I4def}), there are eight terms in \( I_4 \) which contain \( E \): \( S_{E} \), \( S_{AE} \), \( S_{BE} \), \( S_{CE} \), \( S_{ABE} \), \( S_{BCE} \), \( S_{ACE} \), and \( S_{ABCE} \), respectively. Each of those RT surfaces contains the semicircles in those figures and are labeled on those semicircles. Red and blue represent the positivity of the term in \( I_4 \), and the color of the circle represents the positivity of the summation of all RT surfaces labeled on this circle (black means that the area of this circle cancels out in \( I_4 \)). The black arrows below the endpoints of gaps between \( E_{i1} \) and \( E_{i2} \) represent the direction of the modification of those endpoints that enlarges the \( I_4 \). The dashed arrows between each diagram represent the direction of increasing \( I_4 \).

Let us analyze these cases one by one.

The first case is when \( E_i \) connects with \( A \) or \( B \) or \( C \) in all entanglement wedges except \( EW(E) \). As shown in diagram (1.1), splitting \( E_i \) into \( E_{i1} \) and \( E_{i2} \) with a gap between them will decrease \( I_4 \), as one should decrease the length of the gap to make the length of the blue curve decrease and the length of red curves increase in order to increase \( I_4 \). The second case is that \( E_i \) only disconnects with \( A \) in entanglement wedge \( AE \) while connecting with \( A \) or \( B \) or \( C \) in all other entanglement wedges, as shown in diagram (1.2). In this case, moving the endpoints of the gap between \( E_{i1} \) and \( E_{i2} \) will not modify \( I_4 \), so we can enlarge the gap between \( E_{i1} \) and \( E_{i2} \) until the phase transition of \( EW(BE) \) occurs to \( E_{i1} \) and \( E_{i2} \) simultaneously and make the entanglement wedge \( BE \) disconnected, which is shown in diagram (1.3). We can split \( E_i \), which disconnects with \( A \) and \( B \) in the entanglement wedges of \( AE \) and \( BE \) while connecting with other entanglement wedges, again. This time, splitting will increase \( I_4 \) to diagram (2.2) or diagram (3.1). Splitting again, diagram (2.2) or diagram (3.1) will lead to diagram (3.2). Splitting diagram (3.2) will lead to diagram (4.1). However, splitting diagram (4.1) will decrease \( I_4 \). As a result, the maximum of \( I_4 \) might be chosen as the configuration between diagram (4.1) and (4.2), which is the phase transition point. In this case, \( E \) disconnects with \( A \), \( B \), and \( C \) in the entanglement wedges of \( ABE \), \( BCE \), and \( ACE \), which satisfies the disconnectivity condition.

From the above argument, we can see that \( I_4 \) in the disconnected configuration reaches the maximum among all diagrams except diagram (1.1). So we are one step closer to proving the disconnectivity condition. The next step is to rule out the possibility of diagram (1.1) being the diagram with the maximum \( I_4 \). However, this task seems difficult. Instead, we have 
\begin{equation}\label{I4min}
    I_4(E:A:B:C) = 2I_3(A:B:C) - I_4(F:A:B:C),
\end{equation}
where $F=(ABCE)^c$ is the complement of region $ABCE$. We can find that when $ABC$ are fixed regions, if $I_4(F:A:B:C)$ reaches the upper bound, $I_4(E:A:B:C)$ must reach the lower bound.
As a result, we could try to find the configuration of $E$ with the minimal value of \( I_4 \). Then, according to equation (\ref{I4min}), \( F= (ABCE)^c \) will make $I_4(A:B:C:F)$ reach the upper bound.

The procedure of finding the minimal value is exactly the opposite of finding the maximum value. One just has to reverse all arrows in Figure \ref{DISI4}. As we already analyzed before, it is easy to find that there are two configurations which might reach the minimal value: diagram (1.2) and diagram (4.2). Let us analyze diagram (4.2) first. One should enlarge the gap between \( E_{i1} \) and \( E_{i2} \) in order to decrease \( I_4 \). At last, \( EW(ABCE) \) could be disconnected and \( I_4 \) reaches its minimal value. However, in this case, intervals \( E_{i1} \) and \( E_{i2} \) disconnect with all entanglement wedges. We have argued that this case will lead to the result that eliminating \( E_{i1} \) and \( E_{i2} \) will not affect the CMI. This argument is still valid in the \( I_4 \) case. Until now, we understand why diagram (4.2) will reach the local minimal value of \( I_4 \), because it is zero, and the real lower bound must be negative. So, only diagram (1.2) has the right to become the configuration with minimal \( I_4 \). As adjusting the length of the gap will not change \( I_4 \), we can shorten it until a phase transition between diagram (1.1) and (1.2) happens, in which case, \( E \) connects with \( A \) or \( B \) or \( C \) in all entanglement wedges. In this diagram, we mark \( F \) as the region that purifies \( ABCE \); then \( F \) is the collection of gaps between \( E_i \) and \( E_{i+1} \), and the connectivity of \( AE \), \( BE \), and \( CE \) is equivalent to the disconnectivity of \( BCF \), \( ACF \), and \( ABF \), respectively. From equation (\ref{I4min}), \( F \) must be the region that maximizes \( I_4 \), and the disconnectivity condition (\ref{I4disconnect}) is proven.
\end{document}